\documentclass[fleqn,PRC,twocolumn,superscriptaddress,showpacs,preprintnumbers,amsmath,amssymb]{revtex4}
\usepackage{graphicx}% Include figure files
\usepackage{dcolumn}% Align table columns on decimal point
\usepackage{bm}% bold math
\usepackage{color}

\begin{document}

\title{Shear viscosity to entropy
density ratio in the Boltzmann-Uehling-Uhlenbeck model}

\author{ S. X. Li} \affiliation{Shanghai Institute of Applied Physics, Chinese
Academy of Sciences, Shanghai 201800, China} \affiliation{Graduate
School of the Chinese Academy of Sciences, Beijing 100080, China}

\author{D. Q. Fang}\affiliation{Shanghai Institute of Applied Physics, Chinese
Academy of Sciences, Shanghai 201800, China}

\author{Y. G.  Ma\footnote{Author to whom all correspondence should be addressed:
ygma@sinap.ac.cn}}\affiliation{Shanghai Institute of Applied
Physics, Chinese Academy of Sciences, Shanghai 201800, China}

\author{C. L. Zhou}
\affiliation{Shanghai Institute of Applied Physics, Chinese
Academy of Sciences, Shanghai 201800, China}
\affiliation{Graduate
School of the Chinese Academy of Sciences, Beijing 100080, China}

\date{\today}

\begin{abstract}

The ratio of shear viscosity ($\eta$) to entropy density ($s$) for
an equilibrated system is investigated in intermediate energy
heavy ion collisions below 100$A$ MeV within the framework of the
Boltzmann-Uehling-Uhlenbeck (BUU) model .  After the collision
system almost reaches a local equilibration, the temperature,
pressure and energy density  are obtained from the phase space
information and {$\eta/s$} is calculated using the Green-Kubo
formulas. The results show that {$\eta$}/$s$ decreases with
incident energy and tend towards a smaller value around 0.5, which
is not so drastically different from the BNL Relativistic Heavy
Ion Collider results in the present model.

\end{abstract}

\pacs{25.70.-z, 66.20.Cy}

\maketitle

\section{\label{sec:level1}INTRODUCTION}
Studying the behavior of nuclear matter under extreme conditions
is one of the most important problems in heavy ion collisions. Due
to van der Waals nature of the nucleon-nucleon interaction, it is
expected that multifragmentation may exhibit features of
liquid-gas phase transition (LGPT) in intermediate energy
heavy-ion collisions \cite{lab1,lab2}. Evidences of this have been
provided from various observables, such as the nuclear caloric
curve, fluctuation, fragment mass distribution and moment analysis
etc \cite{lab3,lab4,Poch,Nato,Ma-TAMU,Ma-EPJA,Ma-PRL,Gul-EPJA}.
Recent progress on nuclear liquid-gas phase transition has been
reviewed, especially for the signals of LGPT in theory and
experiment \cite{Borderie,WCI}.

Empirical observation of the temperature or incident energy
dependence of the shear viscosity to entropy density ratio
($\eta/s$) for H$_{2}$O, He and Ne$_{2}$ exhibits a minimum in the
vicinity of the critical point for phase transition \cite{lab5}.
Furthermore, a lower bound of $\eta/s>1/4\pi$ obtained by
Kovtun-Son-Starinets (KSS) in certain gauge theories is speculated
to be valid for several substances in nature \cite{lab6,lab7}. In
ultra-relativistic heavy ion collision
\cite{lab8,lab9,Chen1,Kapu,Maj,XuZhe}, people used the shear
viscosity to entropy density ratio to study the Quark-Gluon Plasma
phase and get the minimum value of {$\eta$}/$s$, so it is very
interesting to study shear viscosity or {$\eta$}/$s$ in
intermediate energy heavy ion collision
\cite{Chen2,lab21,Shi,lab11,Shlomo}. Unlike the studies on
$\eta/s$ at relativistic energies, there is still very limited
investigations in intermediate energy heavy-ion collisions.

In this work, we study the thermodynamic and transport properties
of nuclear reaction and try to see how the {$\eta$}/$s$ evolves
with the beam energy or temperature in a transport model. We study
the equilibration of nuclear system within a finite volume using
BUU model. In order to make the system contain enough number of
nucleons in the fixed spherical volume, we choose Au + Au system
in head-on collision (b = 0fm). The system evolves with time for
long enough time so that it is in freeze-out stage.

In the final stage of the central collisions, the system can be
viewed as locally equilibrated.  The equilibrium in intermediate
energy heavy ion collisions can be judged by using the temperature
and other dynamical variables  \cite{lab10}. After the system is
in equilibrium, we calculate the thermodynamic parameters
(pressure, energy density and entropy density) from phase space
information of the system. Shear viscosity coefficient is
calculated from stress tensor fluctuations around the equilibrium
state using Green-Kubo formula \cite{lab11,lab12}. Finally, we
compare {$\eta$}/$s$ in different incident energy with different
nuclear equation of state and  discuss the results.

The rest of the paper is organized as follows: In Sec. II, we
describe the situation of system equilibrium. In Sec. III, we
calculate viscosity coefficient and entropy density. Finally, a
brief summary and  outlook is made in Sec. IV.

\section{\label{sec:level2} equilibration of finite nucleon system}

We calculate the shear viscosity to entropy density ratio $\eta/s$
of an equilibrated nuclear system in intermediate energy heavy ion
collisions using BUU model, which is a one-body microscopic
transport model based upon the Boltzmann equation
\cite{lab13,lab14}.

The BUU equation reads \cite{Bauer}

\begin{align}
      & \frac{\partial f}{\partial t}+ v \cdot \nabla_r f - \nabla_r U
\cdot \nabla_p f  = \frac{4}{(2\pi)^3} \int d^3p_2 d^3p_3 d\Omega
\nonumber
\\ & \frac{d\sigma_{NN}}{d\Omega}V_{12}
 \times [f_3 f_4(1-f)(1-f_2) - f f_2(1-f_3)(1-f_4)] \nonumber
\\ & \delta^3(p+p_2-p_3-p_4).  \label{BUU}
                   \end{align}

It is solved with the method of Bertsch and Das Gupta
\cite{Bertsch}. In Eq.(~\ref{BUU}), $\frac{d\sigma_{NN}}{d\Omega}$
and $V_{12}$ are in-medium nucleon-nucleon cross section and
relative velocity for the colliding nucleons, respectively, and
$U$ is the mean field potential including the isospin-dependent
term:

\begin{equation}
  U(\rho,\tau_{z}) = a(\frac{\rho}{\rho_{0}}) +
  b(\frac{\rho}{\rho_{0}})^{\sigma} + C_{sym} \frac{(\rho_{n} -
    \rho_{p})}{\rho_{0}}\tau_{z},
\end{equation}
where $\rho_0$ is the normal nuclear matter density; $\rho$,
$\rho_n$, and $\rho_p$ are the nucleon, neutron and proton
densities, respectively; $\tau_z$ equals 1 or -1 for neutrons and
protons, respectively; The coefficients $a$, $b$ and $\sigma$ are
parameters for nuclear equation of state.  Two sets of mean field
parameters are used in this work, namely the soft EOS with the
compressibility $K$ of 200 MeV ($a$ = -356 MeV, $b$ = 303 MeV,
$\sigma$ = 7/6), and the hard EOS with $K$ of 380 MeV ($a$ = -124
MeV, $b$ = 70.5 MeV, $\sigma$ = 2). $C_{sym}$ is the symmetry
energy strength due to the density difference of neutrons and
protons in nuclear medium, here $C_{sym} = 32$ MeV is used.

In this work, we focus on the thermodynamic and transport
properties of a nuclear system. For this purpose, we investigate
the process of the head-on Au + Au collision in a spherical volume
with the radius of 5 fm. In the following calculation, we show the
results with the hard EOS except in the case we mentioned.

First we check the evolution of equilibration situation and
temperature. The anisotropy ratio, which is a measure of the
degree of equilibration reached in a heavy-ion reaction, is
defined as£º
\begin{eqnarray}
\label{eq2} R_{p} = \frac{2}{\pi} \frac{R_{\parallel}}{R_{\perp}},
\end{eqnarray}
where $R_{\parallel}=\langle \sqrt{p_x^{2}+p_y^{2}} \rangle$ and
$R_{\perp}=\langle \sqrt{p_z^{2}} \rangle$ are calculated by the
momentum of nucleons in the given sphere. As an example, the time
evolutions of $R_{p}$ for Au+Au systems within a 5fm-radius sphere
at 50 $A$ MeV are shown in Fig. 1(a). When $R_{p}$ approaches to 1
at around 100 fm/c, nuclear system is under equilibrium.

Time evolution of temperature is also used to judge the state of
equilibration. Temperature of the system can be derived from the
momentum fluctuations of particles in the center of mass frame of
the fragmenting source \cite{Aldo}. The variance $\sigma^{2}$ is
obtained from the $Q_{z}$ distribution through
\begin{eqnarray}
\label{eq3} \sigma^{2}=\langle {Q_{z}}^{2} \rangle-\langle Q_{z}
\rangle^{2},
\end{eqnarray}
where $Q_{z}$ is the quadruple moment which is defined by
$Q_{z}=2{p_{z}}^{2}-{p_{x}}^{2}-{p_{y}}^{2}$, and $p_{x},
 p_{y}$ and $p_{z}$ are three components of momentum vector
 extracted from the phase space of
BUU model. If the mean equals zero, the second term vanishes.
$Q_{z}^{2}$ is described by
\begin{eqnarray}
\label{eq4} \langle {Q_{z}}^{2}
\rangle=\int{d^{3}p}{(2{p_{z}}^{2}-{p_{x}}^{2}-{p_{y}}^{2})}^{2}f(p).
\end{eqnarray}
Assuming a Maxwellian distribution for the momentum distribution,
i.e.
\begin{eqnarray}
\label{eq5}
f(p)=\frac{1}{{(2{\pi}{m}T})^{3/2}}e^{-\frac{{p_{x}}^{2}+{p_{y}}^{2}+{p_{z}}^{2}}{2mT}},
\end{eqnarray}
we can obtain
\begin{eqnarray}
\label{eq6} \langle {Q_{z}}^{2} \rangle=4{m}^{2}{A^{2}}T^{2}
\end{eqnarray} after Gaussian integral,
where $m$ is the mass of a nucleon and $A$ is the mass number of
the fragment. For a nucleonic system, we have $A=1$ and can
calculate the evolution of temperature using this equation. Fig.
1(b) shows the temperature's evolution after $25fm/c$, it is seen
that temperature reaches a maximum around 50 fm/c when the system
is in the most compressible stage and then it starts cool down
when the system expands, later on the system tends thermodynamic
equilibrium. For an equilibrated system, the kinetic energy
distributions approach the Boltzmann distribution as time
increases \cite{lab15}. After the expansion process, the system
will approach an equilibrate state, so we can investigate the
viscosity coefficient and entropy density in system.

\begin{figure}
\includegraphics[width=8cm]{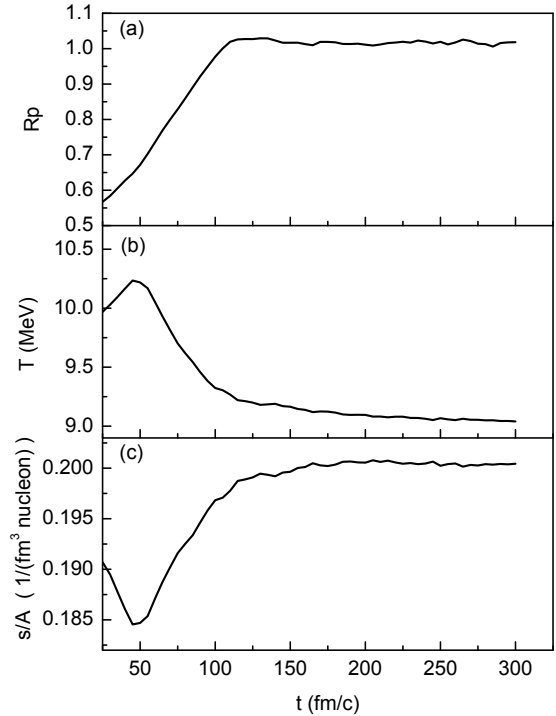}
%\vspace{-0.1truein}
\caption{\footnotesize $R_{p}$ (a), temperature (b) and entropy
density per nucleon (c) as a function of time (after 24 fm/c) for
the head-on Au + Au collision within 5fm-radius sphere at 50 $A$
MeV. }\label{fig1}
\end{figure}

Except temperature, other thermodynamic variables can be
calculated during heavy-ion collisions. Energy density inside a
volume with the 5fm radius can be defined as \begin{equation}
\varepsilon = \frac{1}{V} \sum_{r_i<r_0} E_i,
\end{equation}
where $E_i$ is $\sqrt{p_i^2+m_i^2}$,  $r_i$ is the position of the
$i$-th nucleon in the center of mass and $r_0$ is the selected
radius (here we set $r_0$ = 5 fm) and pressure can be defined as
\begin{equation}
P = \frac{1}{3V} \sum_{r_i<r_0}\frac{p_i^2}{E_i}.
\end{equation}
After we get the energy density,  pressure and temperature,
entropy density can be calculated by the Gibbs formula
\begin{equation}
s = \frac{\varepsilon + P - \mu_n \rho}{T},
\end{equation}
where $\mu_n$ is the nucleon chemical potential and $\rho$ is
nucleon density of system within the given sphere. In principal,
once we have the temperature $T$ and $f(p)$, we can fit to a
Fermi-Dirac function  to extract the chemical potential. However,
we can assume, for simplicity, zero nucleon chemical potential, or
$\mu_n$ can be taken around 20 MeV in the present calculation
\cite{Kono}. In the following calculations of entropy and
$\eta/s$, we only show the results with $\mu_n$ = 20 MeV.  But we
have also checked the results with zero chemical potential, this
will increase the entropy about 8$\%$ and then lead to the
decreasing of $\eta/s$ about 8$\%$. However, this does not change
our conclusions of this work.  Fig. 1(c) shows the entropy density
per nucleon ($s/A$) evolves with time after 25 fm/c. It seems the
entropy density per nucleon reaches a minimum when the system is
in the most compression stage and rises up in the expansion phase,
it finally reaches to an asymptotic value. From the viewpoint of
phase space, the number of occupied states is most limited in the
high density phase. In the high density phase, Fermi blocking
forces some of the nucleons into higher momentum states but still
in a spatially confined region, which makes the entropy density
the minimum. When the system expands, the Fermi blocking is
reduced, the system cools, but more coordinate volume is occupied.
Therefore it is possible that this leads to a non-isentropic
expansion until the system is in the freeze-out stage.

\begin{figure}
%\vspace{-0.1truein}
\includegraphics[width=8cm]{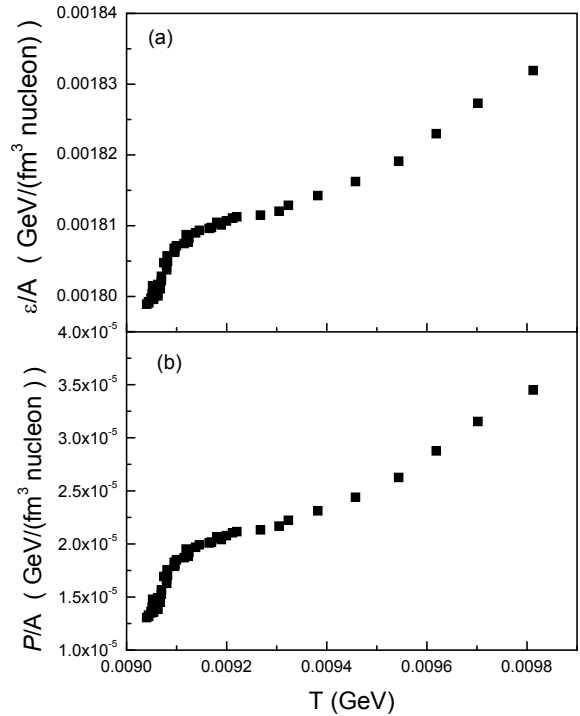}
%\vspace{-0.1truein}
\caption{\footnotesize Energy density per nucleon (a) and pressure
per nucleon  (b) as a function of temperature for the head-on Au +
Au collision within 5fm-radius sphere at 50 $A$ MeV.}\label{fig2}
\end{figure}

Energy density and temperature can be calculated in different time
when the reaction is going on.  Fig. 2(a) shows  energy density
per nucleon versus temperature for the studied system after 25
fm/c. Similarly, pressure per nucleon increases with temperature
is shown in Fig.2(b). From the figure, we can see both energy
density and pressure  increase with temperature. This can be
understood that in a given volume, the increasing of temperature
reflects stronger thermal motion of nucleons, therefore the
kinetic energy will give more contribution on energy density and
pressure.

\section{\label{sec:level3}  viscosity coefficient and entropy density}

Now let's move on the investigation on transport coefficient of
Au+Au system in  a given volume by the BUU model. Viscosity is one
of the transport coefficients which characterize the dynamical
fluctuation of dissipative fluxes in a medium. Transport
coefficients can be measured, as in the case of condensed matter
applications. Also, they should be  in principle calculable from
the first principle. Monte-Carlo simulation for transport
coefficients is a powerful tool when studying transport
coefficients using Green-Kubo relations \cite{lab16,lab17}. In
high energy heavy-ion collisions, calculation of transport
coefficients of shear viscosity for a binary mixture \cite{lab18},
and the calculation of coefficient of a hadrons gas has been
studied \cite{lab19,lab15}. The situation of nuclear gas in
intermediate energy heavy ion collisions is similar to hadrons
gas. To study the extended irreversible dynamic processes, we  use
the Kubo fluctuation theory to extract transport coefficients
\cite{lab12}. The formula relates linear transport coefficients to
near-equilibrium correlations of dissipative fluxes and treats
dissipative fluxes as perturbations to local thermal equilibrium.
The Green-Kubo formula for shear viscosity is defined by
\begin{eqnarray}
\label{eq7} \eta = \frac{1}{T}
\int{d^{3}r}\int_{0}^{\;\infty}{dt}\langle
{\pi_{ij}}(0,0){\pi_{ij}}(\vec r,t) \rangle,
\end{eqnarray}
where $T$ is the equilibrium temperature of the system, $t$ is the
post-equilibration time (the above formula defines $t=0$ as the
time the system equilibrates and is determined by equilibrium
time), and $\langle {\pi_{ij}}(0,0){\pi_{ij}}(\vec r,t) \rangle$
is the shear component of the energy momentum tensor. The
expression for the energy momentum tensor is defined by
$\pi_{ij}=T_{ij}-\frac{1}{3}{\delta}_{ij}T_{i}^{i}$, the momentum
tensor reads \cite{lab15}
\begin{eqnarray}
\label{eq8}
T_{ij}(r,t)=\int{d^{3}p}\frac{p^{i}p^{j}}{p^{0}}f(x,p,t),
\end{eqnarray}
where $f(x,p,t)$ is the phase space density of the particles in
the system. In order to compute an integral, we assume that
nucleons are uniformly distributed in the space. Meanwhile, the
isolated spherical volume with the radius of 5 fm is fixed, so the
viscosity becomes
\begin{eqnarray}
\label{eq9} \eta = \frac{V}{T} \langle {\pi_{ij}(0)}^{2}
\rangle{\tau}_{\pi},
\end{eqnarray}
where $\tau_{\pi}$ is calculated by \begin{equation} \langle
{\pi_{ij}(0)}{\pi_{ij}(t)} \rangle \propto
\exp{(-\frac{1}{\tau_{\pi}})} \label{rex}.
\end{equation}
As shown in Fig. 3(a),   $\langle {\pi_{ij}}(0,0){\pi_{ij}}(\vec
r,t) \rangle$ is plotted as a function of time for Au + Au at 50
$A$ MeV. The correlation function is damped exponentially with
time and can be  fitted by the Eq.~(\ref{rex}) to extract the
inverse slope corresponds as the relaxation time.  Fig. 3(b)
summarizes the relaxation time decrease as the increase of
incident energy, indicating that the system can approach to
equilibration faster at higher incident energy.

\begin{figure}
%\vspace{-0.1truein}
\includegraphics[width=8cm]{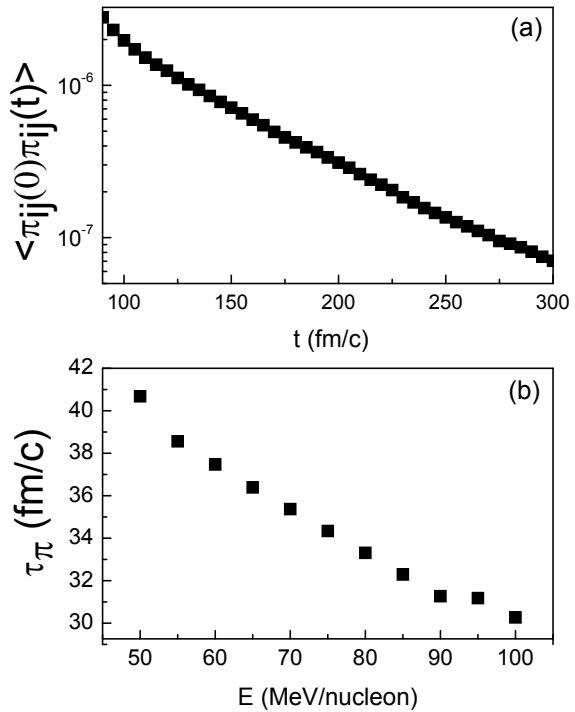}
%\vspace{-0.1truein}
\caption{\footnotesize  (a) $\langle
{\pi_{ij}}(0,0){\pi_{ij}}(\vec r,t) \rangle$ evolves with time for
the head-on Au+Au collision in a given 5fm-radius volume at 50 $A$
MeV; (b) Relaxation time  as a function of incident energy for the
head-on Au+Au collision in a given 5fm-radius volume.
}\label{fig3}
\end{figure}

\begin{figure}
\vspace{-0.6truein}
\includegraphics[width=8cm]{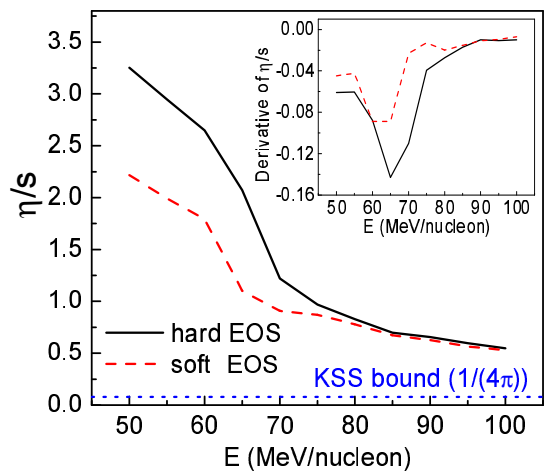}
\vspace{-0.8truein} \caption{\footnotesize (Color online) $\eta/s$
as a  function of beam energy for the head-on Au + Au collision in
a spherical volume with radius of 5 fm. The inset shows the
derivative of $\eta/s$ versus beam energy.}\label{fig4}
\end{figure}

Using the above method, we present the value of {$\eta$}/$s$ as a
function of incident energy  after the studied system has been in
equilibrium in Fig. 4.  The two sets of nuclear equation of state
are used.  The $\eta/s$ value shows a rapid fall as the increasing
of incident energy up to $E < 70 A$ MeV and then drops slowly to a
value close 0.5 when $E > 70 A$ MeV . Since the BUU equation is
one-body theory, fragmentation which originates from the
fluctuation and correlation can not be treated in the present
model. In this case, the phase transition behavior cannot be
predicted in the BUU model. The continuous drop of the ratio of
shear viscosity to entropy density does not show a minimum at a
certain beam energy, which indicates no obvious phase change or
critical behavior in the present model. This is a shortcoming of
the BUU model itself, especially when it is applied to higher beam
energy. Actually, when we calculate the differential values of
$\eta/s$ versus the beam energy (see the inset of Fig. 4), it
seems a turning point around $E \sim$ 65$A$ MeV. This turning
point could indicate the change of dynamical behavior of system,
in other words,  other mechanisms may be needed to be taken into
account in the model especially at higher beam energies, eg.
multifragmentation \cite{Ma95}. Alternatively, lack of minimum
$\eta/s$ perhaps shows that a behavior with a local minimum of
$\eta/s$ at phase transition temperature might not be universal
\cite{ChenJW}. In the present BUU calculation, all calculated
values of $\eta/s$ are well above the conjectured KSS lower bound
of $1/4\pi$ \cite{lab6,lab7}. Comparing these values of $\eta/s$
for our finite nuclei in BUU model,  we see they are not
drastically different either from the RHIC results
\cite{lab9,Song} or from the results of the usual finite nuclei at
low temperature from the widths of giant vibrational states in
nuclei \cite{Shlomo}. As pointed in Ref.~\cite{Shlomo}, it is
possible that the strong fluidity is a characteristic feature of
the strong interaction of the many-body nuclear systems in general
and not just of the state created in the relativistic collisions.
Another interesting point from Fig. 4 is that $\eta/s$ shows EOS
sensitivity in lower beam energy: hard EOS displays larger
$\eta/s$ than the soft one, i.e. larger compressibility of nuclear
matter can lead to higher $\eta/s$ value.

\section{\label{sec:level5}SUMMARY}

In summary, we studied thermodynamic variables as well as
viscosity and entropy density for heavy ion collision after the
system tends towards equilibrium in intermediate energy heavy ion
collisions in the framework of BUU model. The Green-Kubo relation
has been applied for the nucleonic matter in a central region with
a moderate volume when the system has been in equilibrium stage
for central heavy-ion collisions of Au + Au. It is found that the
ratio of shear viscosity to entropy density $\eta/s$ decreases
very quickly before 70$A$ MeV and then drops slowly towards a
smaller value of $\eta/s$ around 0.5 at higher beam energy in
Au+Au system. The  $\eta/s$  are not drastically different either
from the RHIC results  or from the results of the usual finite
nuclei at low temperature. However, no obvious minimum $\eta/s$
value occurs within the investigated energy range. This may
reflect that no liquid-gas phase transition behavior is displayed
in the present model due to the shortcoming of the model itself
which lacks dynamical fluctuation and correlation. Relating the
shortcoming, the equilibrium temperature could be a little higher
than other model (QMD, SMM) which considers fragment formation
\cite{lab20} and while the shear viscosity and entropy density
might be influenced by the cluster formation. Therefore, other
models which can incorporate liquid gas phase transition should be
checked for the shear viscosity and entropy density. For instance,
it will be very interesting to use quantum molecular dynamics-type
model to check if a minimum of $\eta/s$ will occur around
liquid-gas phase transition. The work along this direction is in
progress. Of course, experimental studies on shear viscosity is
more important to demonstrate the relation of {$\eta$}/$s$ and
liquid-gas transition point.

\bigskip
\bigskip

%\section{ACKNOWLEDGEMENTS}
Acknowledgements: Authors acknowledge Prof. W. Zajc for helpful
communications.  This work is supported partially by the National
Natural Science Foundation of China under contract No. 11035009,
No. 10979074, No. 10775168 and No. 10975174,
 Major State Basic Research Development Program in China
under contract No. 2007CB815004, Shanghai Development Foundation
for Science and Technology under contract No. 09JC1416800, and
Knowledge Innovation Project of CAS under Grant No. KJCX2-EW-N01.

\end{document}